\documentclass[
 preprint, amsmath,amssymb,aps,]{revtex4-1}

\usepackage{graphicx}
\usepackage{dcolumn}
\usepackage{bm}
\usepackage{epstopdf}
\usepackage{subcaption}
\usepackage{floatrow}
\usepackage{color}
\newfloatcommand{capbtabbox}{table}[][\FBwidth]
\usepackage[T1]{fontenc}
\usepackage[font=small,labelfont=bf,tableposition=top]{caption}
\usepackage{color}

\newcommand{\Beta}{\textrm{Beta}}
\newcommand{\Bin}{\textrm{Bin}}
\newcommand{\Var}{\textrm{Var}}

\newcommand{\tss}[1]{\textsuperscript{#1}}

\DeclareCaptionLabelFormat{andtable}{#1~#2  \&  \tablename~\thetable}

\begin{document}

\title{The gallium anomaly reassessed using a Bayesian approach}

\author{Joel Kostensalo}%
\email{joel.j.kostensalo@student.jyu.fi}  
\affiliation{Department of Physics, University of Jyvaskyla, Finland}

\author{Santtu Tikka}
\email{santtu.tikka@jyu.fi}
 \affiliation{Department of Mathematics and Statistics, University of Jyvaskyla, 
Finland}

\author{Jouni Suhonen }%
 \email{jouni.t.suhonen@jyu.fi} 
\affiliation{Department of Physics, University of Jyvaskyla, Finland}





\date{\today}
\begin{abstract}
The solar-neutrino detectors GALLEX \cite{Anselmann1995,Hampel1998,Kaether2010}
and SAGE \cite{Abdurashitov} were calibrated by electron-neutrino
flux from the $^{37}$Ar and $^{51}$Cr calibration sources. A deficit in the measured
neutrino flux was recorded by counting the number of neutrino-induced
conversions of the $^{71}$Ga nuclei to $^{71}$Ge nuclei. This deficit was coined
``gallium anomaly'' and it has lead to speculations about beyond-the-standard-model
physics in the form of eV-mass sterile neutrinos. Notably, this anomaly has already
defied final solution for more than 20 years. Here we reassess the statistical
significance of this anomaly and improve the related statistical approaches by 
treating the neutrino experiments as repeated Bernoulli trials
taking into account the fact that the number of the detected $^{71}$Ge nuclei is 
quite small, thus calling for a Bayesian statistical approach. In addition, we take 
into account the systematic errors of the experiments, their correlations, 
theoretical uncertainties and the number of background
solar-neutrino events as a Poisson-distributed random variable. To compare with
the previously reported statistical significances of the anomaly we convert 
the posterior intervals of our Bayesian approach to standard deviations $\sigma$ of
the frequentist approach. We find that our approach reduces the statistical 
significance of the anomaly by $0.8\,\sigma$ for all the adopted theoretical approaches.
This renders the gallium anomaly a statistically weakly supported concept. 
Furthermore, the implications of our 
approach go far beyond the gallium anomaly since the results of many rare-events 
experiments should be reassessed for their limited number of recorded events. 
\end{abstract}
\pacs{Valid PACS appear here}
											

\maketitle



\section*{Main}

The Gallium-based solar-neutrino detectors of the 
GALLEX \cite{Anselmann1995,Hampel1998,Kaether2010} and 
SAGE \cite{Abdurashitov} experiments have been subjected to detection-efficiency 
testing with strong $^{37}$Ar and $^{51}$Cr neutrino sources. The neutrinos emitted 
by these sources have discrete energies below 1 MeV. The detection is based on 
the charged-current neutrino-nucleus scattering reaction
\begin{equation}
\nu_e + \,^{71}{\rm Ga}(3/2^-_{\rm g.s.}) \ \rightarrow \, ^{71}{\rm Ge}(J^{\pi}) + e^-, 
\label{eq:CC-71}
\end{equation}
leading to low-lying states of multipolarity $J^{\pi}$ in $^{71}\rm Ge$, mainly 
to the ground state and the excited states at 175 keV and 500 keV.

The neutrino-nucleus scattering cross section for the scattering to
the ground state can be deduced from the half-life of $^{71}$Ge and is thus well 
known \cite{Bahcall1997}. However, the total cross section contains also the
scatterings to the excited states as well. Being short-lived states, the cross section 
for these states cannot be determined via $\beta$-decay half-life measurements, and 
so other methods must be employed. Two ways for dealing with this have been 
proposed: Use either charge-exchange reactions to probe the Gamow-Teller strength 
for transitions from the ground state of $^{71}\rm Ga$ to the states in $^{71}\rm Ge$
\cite{Frekers:2011zz} or use a microscopic nuclear model, 
such as the nuclear shell model \cite{Haxton:1998uc,Kostensalo2019}, to
directly compute the cross sections for the scattering transitions to
the $^{71}\rm Ge$ final states. With both of these approaches the theoretical 
estimates have been systematically larger than the values reported by GALLEX 
and SAGE, the experimental values being for example 0.87 $\pm$ 0.05 times
the cross sections based on the theory estimates by 
Bahcall \cite{Bahcall1997}. The origins of these discrepancies have been 
previously discussed in \cite{Giunti2011,Haxton:1998uc,Giunti:2012tn}. 
The mismatch between the measured and theoretical cross sections constitutes
the so-called ``gallium anomaly''. 

One of the suggested explanations for the anomaly is the oscillation of the 
electron neutrino to an eV mass-scale sterile neutrino \cite{Giunti2011,Giunti:2012tn}, 
which could potentially also explain the so-called ``reactor-antineutrino anomaly'' 
\cite{Mueller2011,Mention:2011rk,Huber:2011wv}. However, it should be remarked here 
that there is no accepted sterile-neutrino model which could explain the experimental 
anomalies consistently. Less exotic solutions to the
reactor-antineutrino anomaly, like the proper
inclusion of first-forbidden $\beta$-decay branches in the construction of the
cumulative antineutrino spectra, have also been suggested \cite{Hayen2019a,Hayen2019b}.


The GALLEX and SAGE experiments (gallium experiments for short)
can be thought of as repeated \emph{Bernoulli trials} 
with a single $^{71}$Ga atom in a small area (measured in cm$^2$) with a projectile 
neutrino hitting the square in a uniformly distributed random spot. The result of 
one of these trials can be either a ``success'' (i.e. neutrino-nucleus scattering 
happens) or a ``failure'' (no scattering). The cross section (in cm$^2$) is the 
probability of the scattering to occur. The experiments record the number of 
germanium atoms produced (from which one can deduce the number of events), the 
neutrino flux, and the number of target atoms to some finite accuracy, reporting 
the cross section as a (possibly asymmetric) normal distribution. In the 
papers \cite{Anselmann1995,Hampel1998,Kaether2010,Abdurashitov} the statistical 
errors are related to the number of Bernoulli trials (neutrino flux, number of 
target nuclei) and successes (number of $^{71}$Ge atoms produced). While the 
normal-distribution approximation is asymptotically valid for large number of
trials (and events), this condition 
is not well satisfied in the gallium experiments (or in any rare-events experiment
with a small number of recorded events), 
since the number of observed events is small.   

The small number of events recorded in the gallium experiments produces a major 
source of uncertainty in their reported results. The 
number of events in the experiments varied between approximately 360 and 
520 \cite{Anselmann1995,Hampel1998,Kaether2010,Abdurashitov,GallexPC}. Basic 
probability theory tells us that given the number of successes $s$ and attempts $n$ in 
a repeated Bernoulli trial the success probability $p$ has a likelihood 
function ${\rm Beta}(s,n-s)$. The relative error (the ratio of the standard deviation 
to the expected value) for small success probabilities follows the law $1/\sqrt{s}$, 
and is thus valid for the neutrino experiments. This uncertainty is then 4.2--5.2 \% 
in the gallium experiments. Moreover, the cross-section distribution is highly 
asymmetric for such small success probabilities. Since the relative uncertainty 
is not proportional to the number of trials but only depends on the number of 
successes, we can take the small area which the neutron hits to be 1 cm$^2$, 
since then the numerical value of the cross section is the probability of success 
in a single trial. Note that we could have equivalently picked an area 
e.g. 0.25 cm$^2$ with the number of trials being reduced to fourth and the 
parameter here being four times the cross section and the results would remain 
unchanged.

In the analysis of the GALLEX and SAGE experiments the production rate of 
$^{71}$Ge, that is, how many neutrinos interacted with the detector in a day, 
was assessed for the individual runs (3--28 days) separately. The analysis was 
based on a maximum likelihood fit with constant solar neutrino background. 
The final result was derived by taking an average weighted by the inverse 
variances of the individual runs. However, there is a much more simple way 
to deal with the solar-neutrino background and to combine the individual runs. 
Since we assume that the neutrino interactions are independent we can see that the 
events are \emph{exchangeable} \cite{gelman_bda}. This means that we do not get 
any more information by knowing the number of events in the individual runs. 
The total number of events is a \emph{sufficient statistic} 
(see e.g. \cite{davison}) meaning that all the relevant information regarding 
the probability of success (cross section) is included in this number. Thus 
a more straight-forward way would be to measure the total number of events 
(including the background) and subtract the number of background events as a 
Poisson distributed variable. The resulting distribution can then by 
calculated by simulation.


In this paper we revisit the statistical significance of the gallium anomaly
by reassessing the ratio $R$ between the cross sections of the GALLEX and SAGE 
experiments and those of the theory using a Bayesian approach. In this spirit 
we construct posterior distributions for $R$ and our simulation-based approach allows 
us to take into account properly such details as the correlations between 
the systematic errors of the two GALLEX experiments as well as between the 
two SAGE experiments.


	\begin{figure}
\includegraphics[width=1.0\textwidth]{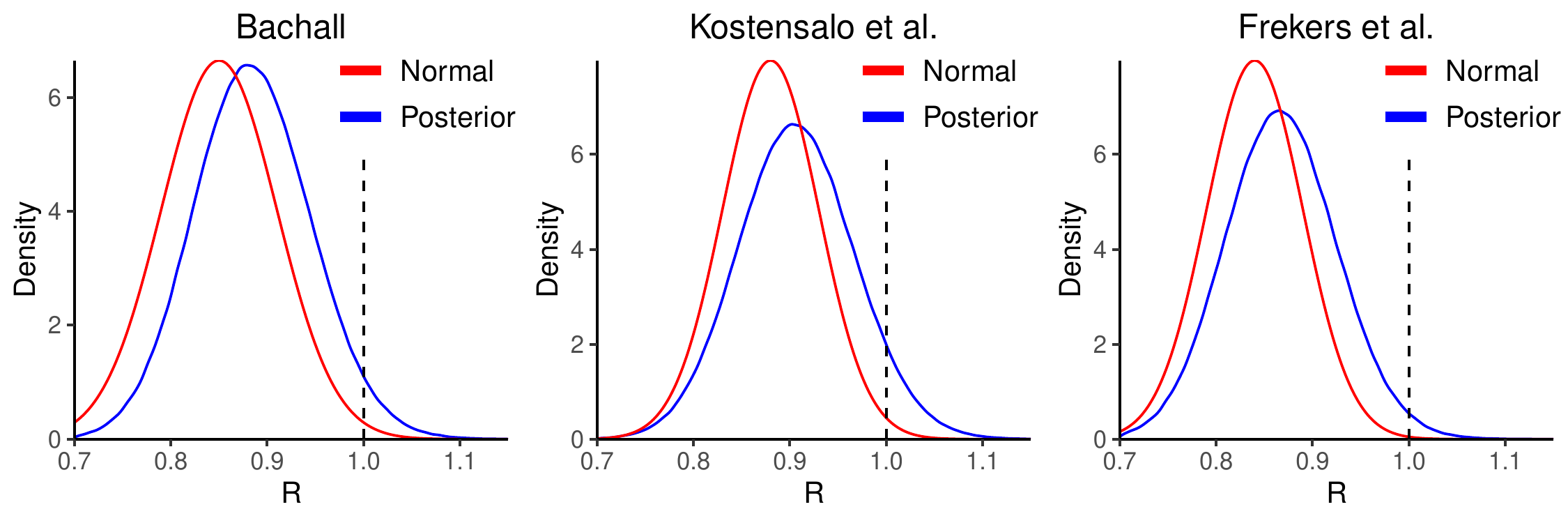}
\caption{Posterior distributions for the experiment-to-theory ratio $R$. The 
normal distributions are those used in the previous analyses of the gallium experiments.}
\label{fig:posteriors}
\end{figure}

The difference between the simple normal distributions used in 
the recent survey \cite{Kostensalo2019} 
and the properly constructed posterior distributions for the ratio $R$ are shown 
in Fig.~\ref{fig:posteriors} for the estimates of Bahcall \cite{Bahcall1997}, the 
shell-model results of Kostensalo et al. \cite{Kostensalo2019} and the charge-exchange 
results of Frekers et al. \cite{Frekers:2011zz}. As clearly visible from the figures, 
the posterior distributions are wider than the previously adopted normal distributions 
and they are shifted slightly to the right. In addition, the skewness of these 
distributions shifts probability to the right of 
$R=1$ relative to the normal distributions. Since the uncertainty related to 
estimating the success probability $p$ of a binomial distribution is proportional to the 
inverse square root of the number of successes, this error is larger in the 
experiments with lower number of successes. This means that for the normal distributions
the experiments with a low number of events are overweighted relative to the 
rest of the experiments, making the anomaly appear larger than it actually is.

\begin{table}
\centering
\begin{tabular}{cccc}
\hline
Theory & Posterior ETI & Significance ($\sigma$) & Normal \cite{Kostensalo2019} ($\sigma$)\\
\hline
              Bachall  &         0.936 &         1.85 &    2.6 \\
        Bachall corr.  &         0.894 &         1.62 &        \\
    Kostensalo et al.  &         0.873 &         1.52 &    2.3 \\
       Frekers et al.  &         0.974 &         2.22 &    3.0 \\
 Frekers et al. corr.  &         0.942 &         1.90 &        \\
\hline
         Combined theory  &         0.915 &         1.72 &        \\
\hline
\end{tabular}
\caption{Magnitude of the gallium anomaly in different theoretical frameworks. 
The column ``Posterior ETI'' gives the width of the smallest equivalently-tailed 
posterior interval in which the experiment to theory ratio 1 is included. 
Column ``Significance'' gives the corresponding significance for a normal distribution 
in standard deviations $\sigma$. The last column gives the previously reported 
standard deviations, all distributions being assumed to be normal 
and the systematic errors independent, even for the same detectors. 
The second and fifth rows include a 30\% tensor 
correction for the excited states. The ``Combined theory'' in the last row 
includes the tensor-corrected Bachall, Kostensalo et al., and the tensor-corrected 
Frekers el al. results weighted by their inverse variances. }
\label{tbl:significance}
\end{table}

The statistical significances of the results are given in Table~\ref{tbl:significance}. 
The results are reported as equivalent-tailed posterior intervals (ETI), that is, the 
null-hypothesis (no anomaly) is included, for example, in the 93.6\% posterior interval 
of the uncorrected Bahcall results. To allow a simple comparison with the previously 
reported results, this is also expressed in a frequentistic way as sigmas of a
normal-distribution (the column ``Significance''). 
The drop in the corresponding frequentistic significance is about 0.8 $\sigma$ for 
all the theoretical models. The most recent theoretical estimate by 
Kostensalo et al. gives no evidence in favor of the gallium anomaly. Furthermore, 
the combined theoretical results also includes $R=1$ in the 95\% ETI. Without any 
tensor corrections the Frekers et al. results still noticeably deviate from the 
GALLEX and SAGE results but only at 2.22 $\sigma$ instead of the previously 
reported 3.0 $\sigma$. Tensor contributions of the magnitude proposed 
in \cite{Kostensalo2019} seem to be able to explain away most of the 
remaining discrepancies.

In this Letter the statistical significance of the so-called ``gallium anomaly'' 
was reassessed using a Bayesian approach, in which we constructed posterior 
probability distributions for the ratio $R$ of the experimental and theoretical cross 
sections. It was pointed out that a neutrino-detection experiment can be 
formulated as a repeated \emph{Bernoulli trial} with the reported cross section 
being the probability of ``success'' i.e. a detected event. This means, that even with 
perfect knowledge of the number of detected events, neutrinos hitting the target, and 
number of target nuclei, one would end up with a beta distribution for the 
cross section. This conclusion goes even beyond the present context and embraces
all rear-events experiments with a limited number of detected events.

The size of the gallium anomaly was assessed by a simulation taking into account 
the systematic and statistical errors in the GALLEX and SAGE experiments, 
theoretical uncertainties, correlations in systematic errors, the number of the
background solar-neutrino events as a Poisson distributed random variable, as well 
as the relatively small number of neutrinos observed. The recent shell-model results of 
Kostensalo et al. were shown to agree with the GALLEX and SAGE results within 
the 95 \% posterior interval. In a previous frequentistic analysis \cite{Giunti2011}, 
this significance was reported as 2.3 $\sigma$. Taking into account all the 
theoretical estimates with proper corrections for tensor contributions, the 
theoretical and experimental results are shown to agree within the 95 \% posterior 
interval. This means that there is little evidence of the gallium anomaly 
relating to new physics.

\section*{Methods}

We adopt a Bayesian approach \cite{gelman_bda} to reassess the significance of the 
discrepancies between the theoretical cross sections and the results from GALLEX 
and SAGE gallium experiments. We construct posterior distributions for the \tss{51}Cr 
and \tss{37}Ar cross sections from which we obtain a posterior distribution for 
the theory-to-experiment ratio $R$. The approach is motivated by a multitude of reasons: 
First, we can easily incorporate various heterogeneous sources of uncertainty in 
the statistical model, including the correlated systematic errors of the two GALLEX 
experiments, without having to resort to approximations. Second, while the models 
for the true number of unobserved events, the total number of neutrinos and the cross 
sections are fairly simple, the hierarchical structure quickly becomes more 
complicated when we connect the actual measured quantities to the true underlying 
latent variables, making a graphical model suitable for the task. Third, powerful 
computational methods are readily available for Markov chain Monte Carlo (MCMC) 
simulations.

\begin{table}
\centering
  \caption{The number of events (neutrino-nucleus interactions) and the number 
of Bernoulli trials in each gallium experiment based on all the available information. 
A single trial consists of one neutrino hitting a 1 cm$^2$ square with a single 
gallium atom in a uniformly distributed random spot. To avoid underestimating 
the thickness of the tails we avoid approximating e.g. the product of normal 
distributions as a normal distribution. The reports of SAGE included more 
details in regard to the experimental set-up, which is why the expression 
for the number of trials differs from GALLEX. While the normal approximation 
would work fine here we want to stress that this is not true in general.} 
\begin{tabular}{lccc}
	\hline
    & Method & Trials ($10^{44}$) & Events\\
  \hline
GALLEX 1 & $^{51}$Cr & $703.9^{+12.21}_{-17.76}$(stat.) $ ^{+3.520}_{-3.942}$(syst.)  
& $389.76\pm 38.28$  \\
GALLEX 2 & $^{51}$Cr &  $775.6^{+37.04}_{-17.76}$(stat.) $^{+3.878}_{-4.343}$(syst.) 
& $365.93\pm41.82$\\
SAGE 1 & $^{51}$Cr & $6.766\times(72.6\pm0.2)\times(1.9114\pm0.022)^{+5.7\%}_{-5.6\%}$(syst.)  
& $518.21\pm62.93$ \\
SAGE 2 & $^{37}$Ar & $6.603\times(72.6\pm0.2)\times(1.513\pm0.007)^{+5.4\%}_{-5.2\%}$(syst.)  
& $401.58^{+36.51}_{-32.86}$ \\
	\hline
\end{tabular}
\label{tbl:exs}
\end{table}

Let index $i \in \{G_1,G_2,S_1,S_2\}$ denote the experiment. 
The neutrino source used in each experiment is indexed 
by $j(i)$ such that $j(G_1) = j(G_2) = j(S_1) = {}^{51}\textrm{Cr}$ 
and $j(S_2) = {}^{37}\textrm{Ar}$. The unobserved theoretical cross sections to be 
estimated are $\sigma_{j(i)}$, which are then compared to the theoretical cross 
sections $\sigma_{\textrm{th},j(i)}$ in order to access the experiment-to-theory 
ratio $R$. The true number of observed events in the experiments is $y_i$ and the true 
total number of trials is $n_i$. The numbers used in this work are given in 
Table \ref{tbl:exs}. The number of trials includes a small correction, 
between 0.3 and 1.7 \%, in order to reproduce the best estimate for the experimental 
rates. This is due to round-off errors and the limited accuracy of the 
reported run times.  

We construct a hierarchical model for the experiments as follows. 
The repeated independent Bernoulli trials result in a binomial likelihood 
\[
    y_i|n_i,\sigma_{j(i)} \sim \Bin(n_i, \sigma_{j(i)})
\]
for all $i \in \{G_1,G_2,S_1,S_2\}$.
We select highly uninformative prior distributions for the unobserved cross 
sections $\sigma_{j(i)}$ by taking
\[
    \begin{aligned}
        \sigma_{^{51}\textrm{Cr}} &\sim \Beta(1/2, 1) \\
        \sigma_{^{37}\textrm{Ar}} &\sim \Beta(1/2, 1).
    \end{aligned}
\]
The distribution $\Beta(1/2, 1)$ has most of its probability mass centered 
near $0$, reflecting our a priori understanding that the unknown cross sections 
are more likely to be small than large while residing somewhere on the 
interval $(0, 1)$. In essence, we do not make any strong subjective claims 
about the cross sections and choose to rely on the information obtained 
from the experiments.

Due to the conjugacy of the beta and binomial distributions, we obtain 
the posteriors directly as beta distributions with updated parameters
\[
    \begin{aligned}
        \sigma_{^{51}\textrm{Cr}}|y_{G_1},y_{G_2},y_{S_1},n_{G_1},n_{G_2},n_{S_1} 
&\sim \Beta\left(\frac12 + \sum_{i \neq S_2} y_i, 1 + \sum_{i \neq S_2} (n_i - y_i)\right) \\
        \sigma_{^{37}\textrm{Ar}}|y_{S_2},n_{S_2} &\sim \Beta\left(\frac12 + y_{S_2}, 
1 + n_{S_2} - y_{S_2}\right).
    \end{aligned}
\]
The experiment-to-theory ratio can now be expressed as a weighted average
\[
    R = \widetilde{w}_{^{51}\textrm{Cr}} 
\frac{\sigma_{^{51}\textrm{Cr}}}{\sigma_{\textrm{th},{}^{51}\textrm{Cr}}} +
        \widetilde{w}_{^{37}\textrm{Ar}} 
\frac{\sigma_{^{37}\textrm{Cr}}}{\sigma_{\textrm{th},{}^{37}\textrm{Ar}}},
\]
where the weights $\widetilde{w}_{j(i)}$ are the normalized inverse posterior variances
\[
    \begin{aligned}
    \widetilde{w}_{j(i)} &= \frac{w_{j(i)}}{w_{^{51}\textrm{Cr}} + w_{^{37}\textrm{Ar}}}, \\
    w_{^{51}\textrm{Cr}} &= 
\frac1{\Var(\sigma_{^{51}\textrm{Cr}}|y_{G_1},y_{G_2},y_{S_1},n_{G_1},n_{G_2},n_{S_1})}, \\
    w_{^{37}\textrm{Ar}} &= \frac1{\Var(\sigma_{^{37}\textrm{Ar}}|y_{S_2},n_{S_2})}.
    \end{aligned}
\]

If the required quantities $n_i$ and $y_i$ were known exactly, we could simply 
generate values from the posterior distributions of the cross sections 
$\sigma_{j(i)}$ and from the reported distributions for the theoretical cross 
sections $\sigma_{\textrm{th},j(i)}$ to obtain a posterior distribution for $R$. 
However, there is additional uncertainty associated with the measurements in 
each experiment which we take into account in our model. 

We simulate the posterior distributions of the cross sections via MCMC using 
JAGS (Just Another Gibbs Sampler, \cite{plummer_jags}) in the statistical 
software R \cite{Rsoft}. In order to estimate the experiment-to-theory ratio $R$, 
we also simulate values for the theoretical cross sections $\sigma_{\textrm{th},j(i)}$ 
from distributions given in the literature. For every theory of 
table~\ref{tbl:significance}, the simulation is carried out using 10 chains 
with different initial values. One million samples are drawn from each chain 
with a warm-up period of 100~000 samples, while only including every 10th draw 
in the final posterior sample, totaling one million draws from the posterior for each theoretical estimate. Convergence of the MCMC chains was monitored 
using the adjusted potential scale reduction 
factor \cite{gelman_rubin1992,brooks_gelman1998}, which is a suitable criterion 
in this case since the posterior of $R$ is approximately Gaussian. 
This particular simulation is not very computationally demanding and 
can be easily performed with a modern laptop.

\section*{Acknowledgements}

This work has been partially supported by the Academy of Finland under the Academy 
project no. 318043. J. K. acknowledges the financial support from Jenny and Antti
Wihuri Foundation. S. T. was supported by the Academy of Finland under the 
Academy project no. 311877.


\end{document}